# HT-Ring Paxos: Theory of High Throughput State-Machine Replication for Clustered Data Centers

Vinit Kumar[1] and Ajay Agarwal[2]

[1] Associate Professor with the Krishna Engineering College, Ghaziabad, India.
 (Phone: +919971087809; e-mail: vinitbaghel@gmail.com)

[2] Professor with SRM University, DELHI-NCR Campus, Modinagar, Ghaziabad, India.
 (Phone: +919917083437; e-mail: ajay.aagar@gmail.com)

**Abstract**

Implementations of state-machine replication (SMR) prevalently use the variants of Paxos. Some of the recent variants of Paxos like, Ring Paxos, Multi-Ring Paxos, S-Paxos and HT-Paxos achieve significantly high throughput. However, to meet the growing demand of high throughput, we are proposing HT-Ring Paxos, a variant of Paxos that is basically derived from the classical Paxos. Moreover, it also adopts some fundamental concepts of Ring Paxos, S-Paxos and HT-Paxos for increasing throughput. Furthermore, HT-Ring Paxos is best suitable for clustered data centers and achieves comparatively high throughput among all variants of Paxos. However, similar to Ring Paxos, latency of the HT-Ring Paxos is quite high as compared with other variants of Paxos.

# 1. Introduction

Most of the database applications use State Machine Replication (SMR) [1] [6] for increasing the availability and performance of the database systems. Replication of data provides that the failure of one or more replicas does not prevent the operational replicas from executing client requests. Implementations of SMR prevalently use the variants of Paxos [4], such as, Google's Megastore [25], chubby lock service [14] and yahoo's Zab [24] are few popular variants of Paxos.

In leader based protocols, bottleneck is generally found at the leader and the maximum throughput is limited by the leader's resources (such as CPU and network bandwidth). Moreover, increasing the number of client requests results in a decrease of the throughput. Since the bottleneck is at the leader, more additional replicas may not improve performance; in fact, it decreases throughput since the leader requires to process additional messages.

Ring Paxos [23] offloads the leader by ordering of *ids* (of client requests) by the leader instead of client requests, dissemination of requests and learned-*ids* by the leader through ip-multicasting, a ring of acceptors (it reduces the number of messages sent to other acceptors and received from other acceptors by the leader), batching of requests at the leader and use of pipelining (i.e. parallel execution of ring Paxos instances).

However, in Ring Paxos leader still requires to handle all client communications, assigns unique *id* to client requests, disseminates client requests to all acceptors and learners. Any client also requires knowing about the leader; if leader fails then service will interrupt until the election of a new leader.

S-Paxos [29] offloads the leader by distributed client communications among all non-faulty replicas, disseminating client requests among replicas in a distributed manner, ordering of *ids* (of client requests) by the leader instead of client requests using *classical Paxos* [3] [4], batching the client requests and use of pipelining.

However, in S-Paxos, every non-faulty replica (including leader) receives all client requests either directly from clients or through other replicas (all these requests may reach to the leader in less number of messages compare to Ring Paxos, because of the batching at various replicas). Moreover, leader partially disseminates the client requests and partially handles client communications. In addition, leader also handles all messages belongs to classical Paxos. High number of messages at leader adversely affect throughput.

HT-Paxos [33] adopts all aforementioned concepts of S-Paxos for offloading the leader. In addition, it further offloads the leader by eliminating the work of handling client communications and client request dissemination from the leader, instead it only receives the batch *ids* (or request *ids*) and arranges them in an order (unlike S-Paxos and Ring Paxos). Moreover, as compare to S-Paxos, HT-Paxos significantly reduces the acknowledgement messages at disseminators in large clustered data centers (unlike S-Paxos, where every disseminator sends acknowledgement messages to every other disseminator). Therefore, leader as well as other disseminators becomes truly lightweight and hence for any large clustered data center, HT-Paxos provides higher throughput. Furthermore, it uses the concept of multiple LANs for avoiding the collision that adversely affect throughput. In S-Paxos every replica is also a broadcaster. Multiple broadcasters on a single LAN may adversely affect throughput.

In this paper, we are proposing HT-Ring Paxos (HT stands for high throughput) a variant of Paxos, that is basically derived from the classical Paxos and among other high throughput Paxos protocols, it largely adopts the basic fundamental concepts of Ring Paxos. In addition, it also incorporates some fundamental concepts of S-Paxos and HT-Paxos to further offload the leader. HT-Ring Paxos is best suitable for clustered data centers and achieves comparatively high throughput among all variants of Paxos. Moreover, like other high throughput Paxos protocols, HT-Ring Paxos also offloads the leader by ordering of *ids* (of client requests) by the leader instead of client requests and using the concepts of batching and pipeline. Furthermore, HT-Ring Paxos adopts a ring of acceptors (Like Ring Paxos, for reducing the total number of messages sent to other acceptors and received from other acceptors by the leader, it also reduces the bandwidth requirements). However, HT-Ring Paxos adopts distributed client communications among all non-faulty replicas as well as it also disseminates client requests among replicas in a distributed manner (Unlike Ring Paxos). HT-Ring Paxos uses the concept of multicasting for dissemination of requests and learned-*ids* (Like Ring Paxos). In order to avoid collisions (multiple sources of ip-multicast on a single LAN largely impact throughput [19]) and to increase reliability of communication channel, it uses the concept of multiple LANs (Like HT-Paxos).

Organization of this paper is as follows, Next section presents a system model. Moreover, Section 3 proposes the HT-Ring Paxos. While Section 4 presents a comparative analysis of proposed work with other related work. Finally, concluding Section discusses the advantages of HT-Ring Paxos.

## 2. System Model

HT-Ring Paxos is basically derived from the classical Paxos and also uses some fundamental concepts of Ring Paxos, S-Paxos and HT-Paxos. Like classical Paxos, HT-Ring Paxos have the three classes of agents as *proposers*, *acceptors* and *learners*, wherein each *acceptor* also assumes a role of *coordinator*. Among all *coordinators*, one *coordinator* works as a *leader*. HT-Ring Paxos has the same fundamental guarantees as with the case of the classical Paxos, like, *Nontriviality, Stability, Consistency and Liveness*.

We assume that clustered data center have multiple LANs (local area networks), all acceptors and learners subscribe to all the LANs. We further assume that at most and at least one acceptor can multicast or broadcast for each LAN. Moreover, we call such an acceptor as a *broadcaster*. Any acceptor can be a broadcaster of one or more LANs. On being a broadcaster of multiple LANs, acceptor randomly chooses a LAN from all such LANs for broadcasting or multicasting. Furthermore, *leader* will always be a *broadcaster*. We also assume that for any two non-faulty acceptors, the difference of number of LANs for which these are *broadcaster* cannot be more than one. At any computing site, there could be at most one acceptor and one learner. Each computing node has two buffers for each LAN, one for incoming messages and another for outgoing messages.

Like classical Paxos, we assume that agents communicate through messages. These messages can take arbitrarily long time for reaching their destinations, can be duplicated, can be delivered out of order, and can be lost. Moreover, system detects all corrupted messages and considers such messages as lost. Furthermore, agents discard duplicate messages and proposals.

Like classical Paxos, we assume the customary partially synchronous, non-Byzantine, and distributed model of computation. Thereby agents may fail by stopping, may restart, may operate at arbitrary speed, and always perform an action correctly. Agents have access to stable storage whose state survives failures.

We assume that, at least $\lfloor n/2 \rfloor + 1$ acceptors will always remain non-faulty out of the total *n* acceptors and at least one learner will always be non-faulty. Furthermore, like Ring-Paxos, a circular ring of acceptors is assumed which constitutes a majority. Any failure of acceptor in a ring requires a view change.

For sending a message, two primitives are used (i) **Send < *message* >** to one receiver (ii) **Multicast < *message* >** to multiple receivers. *Send* primitive is for one to one communication and *Multicast* primitive represents that sender sends a single message but specified multiple receivers can receive this message. We can implement this multicasting by using Ethernet/hardware multicasting or by using IP multicasting. Moreover, we also assume that if any agent does not get intended reply then it resends the message after a certain *TIMEOUT*.

## 3. HT-Ring Paxos

### 3.1 An Overview

Like classical Paxos, HT-Ring Paxos may also execute the various instances of the protocol at once, where each instance has a unique instance number *i* and learner of each instance can learn only a single value. At various sites, learners of the same instance number will learn the same value. Moreover, all learners learn the values in the order of instance numbers. Unlike classical Paxos, HT-Ring Paxos achieves the consensus on the *id* rather than *request* or *proposal* (like, Ring Paxos, S-Paxos and HT-Paxos). Because, in general, it reduces the bandwidth requirements.

Leader election protocol elects a leader and sets the *leader* variable of leader as *TRUE* and *leader* variable of others as *FALSE*. Moreover, leader election protocol also achieves consensus on *I* and *lsn w*here, *I* is the highest known instance number among all the non-faulty coordinators at the time of leader election (coordinator may read instance numbers written by any agent on the stable storage). We consider its initial value as null. Whereas, *lsn* is a leader sequence number. Initially its value is null and non-faulty coordinators at the time of leader election increments its value by one and then achieve consensus on the maximum value of *lsn* among any majority of non-faulty coordinators. Leader election protocol writes the updated values of *leader*, *I* and *lsn* on stable storage.

Any proposer (client) sends *request* (*request* contains a *proposal* and their unique *id*) to any one coordinator (randomly chosen). Moreover, if proposer (client) does not receive a reply message < *id* > in a reasonably long time, then it periodically sends same *request* to any one coordinator (randomly chosen) until it gets a reply.

After receiving a *request* from any proposer, coordinator inserts the *request* into *req_set,* if it does not exist. After that, if it is a *broadcaster* then it multicasts the *request* to all other coordinators and learners, in addition, on being a leader too, it calls a new instance of this protocol. Otherwise, it forwards the *request* to any randomly chosen broadcaster.

After receiving a *request* from any coordinator through one to one communication, coordinator inserts the *request* into *req_set,* if it does not exist. After that, if it is a broadcaster then it multicasts the *request* to all other coordinators and learners. In addition, on being a leader too, it calls a new instance of this protocol.

As explained in [5], variables $crnd[c], cval[c], rnd[a], vrnd[a], vval[a]$ in HT-Ring Paxos also have the same meanings as in classical Paxos. HT-Ring Paxos updates these variables like classical Paxos but the difference here is that the variables $cval[c], vval[a]$ either have an *id* of any *request* (instead of any *request*) or a null value.

Upon startup, if $i < I$ (highest known instance number), no value has been learned at instance $i$ and $crnd[c] \neq lsn$ then after assigning the value of $lsn$ to $crnd[c]$, leader sends phase 1a messages to any majority of acceptors. Moreover, if $crnd[c] = lsn$ then it sends phase 2a message to the successor.

In HT-Ring Paxos, how the messages of various phases reach their destination is quite different as compared to classical Paxos.

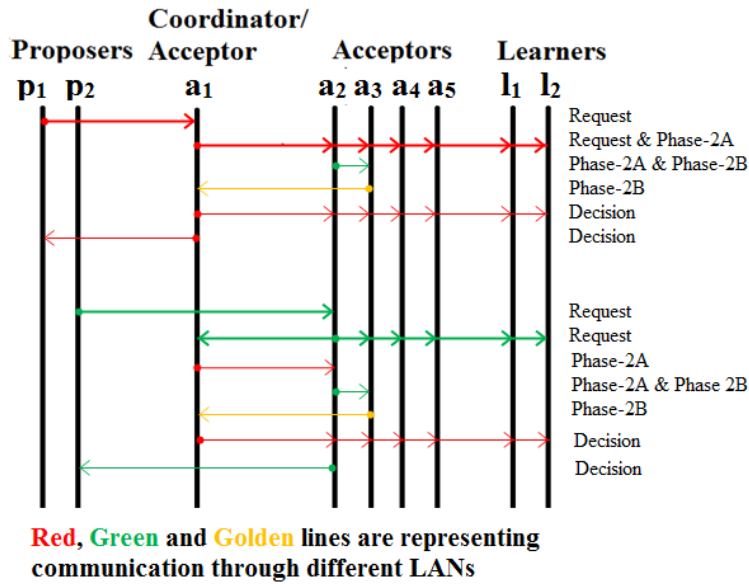

**Fig. 1** Message flow as per the basic algorithm of HT-Ring Paxos under no failures

### 3.2 Pseudo Code of HT-Ring Paxos

---
*Algorithm 1:* **HT-Ring Paxos**
---

1. **Initialize** *leader, I, lsn, broadcaster* from the stable storage /* global variables that can also be used in HT-Ring Paxos (*i*). Leader and broadcaster election algorithms can also modify the values of these variables*/
2. /* Coordinator's Task */
3. **Upon** (startup)
4. **If** $(leader = true)$ **then**

5.     **For** $\forall i : (1 \leq i \leq I \land \text{No value has been learned in instance } i)$ **do**
6.         Call HT-Ring Paxos($i$)
7.     $i \leftarrow I$
8.     **Repeat** step 9 and 10 **if** $(i < |req\_set|)$
9.         $i \leftarrow i + 1$
10.        Call HT-Ring Paxos($i$)
11. **upon** receiving ( *request* ) from any proposer *p*
12.   **If** $(\exists x : (x \in learned\_set) \land (id \in x) \land (id \in request))$ **then** send ($id$) to the proposer *p*
13.   **else if** $(request \notin req\_set)$ **then**
14.       $req\_set \leftarrow request$
15.     **if** $(broadcaster = true)$ **then**
16.       multicast ( *request* ) to all other coordinators and learners
17.     **If** $(leader = true)$ **then**
18.        $i \leftarrow i + 1$
19.        Call HT-Ring Paxos($i$)
20.     **else**
21.      let *br* is a randomly chosen broadcaster
22.      send ( *request* ) to *br*
23. **upon** receiving ( *request* ) from any coordinator *c*
24.   **if** $((broadcaster = true) \land (request \text{ received through one to one communication}))$ **then**
25.     multicast ( *request* ) to all other coordinators and learners
26.   **if** $(request \notin req\_set)$ **then**
27.     $req\_set \leftarrow request$
28.   **If** $(leader = true)$ **then**
29.     $i \leftarrow i + 1$
30.     Call HT-Ring Paxos($i$)
31. **upon** receiving ($id$) from any learner *l*
32.   **if** $\exists request : (request \in req\_set \land id \in request)$ **then** Send ( *request* ) to learner *l*
33. /* *Acceptor's Task* */
34. **upon** receiving ( PHASE 1A, $crnd[c]$, $i$) or (PHASE 2A, $crnd[c]$, $cval[c]$, $i$) or (PHASE 2A & 2B, $crnd[c]$, $cval[c]$, $sn$, $i$) from any coordinator *c*
35.   **If** ( instance $i$ does not exist) **then** Call HT-Ring Paxos($i$)
36. /* *Learner's Task* */
37. **upon** receiving ( *request* ) from any broadcaster     /* if coordinator is not available on this site */
38.   $req\_set \leftarrow request$
39. **upon** receiving (LEARNED, $cval[c]$, $i$) from any coordinator *c*
40.   **If** ( instance $i$ does not exist) **then** Call HT-Ring Paxos($i$)

---

*Algorithm 2:* **HT-Ring Paxos ($i$)**

---

1. **Initialize** $crnd[c], cval[c], rnd[a], vrnd[a], vval[a], id, sn$ to null
2. /* *Task for any coordinator c* */
3. **upon** (startup)

4.  **if** $((leader = true) \wedge (i \leq I))$ **then**
5.      Read $(crnd[c], cval[c])$ from the stable storage for the instance number $i$, if exist.
6.      **if** $(crnd[c] = lsn)$ **then**
7.          send (PHASE 2A, $crnd[c]$, $cval[c]$, $i$) to successor
8.      **else**
9.          $crnd[c] \leftarrow lsn$
10.         send ( PHASE 1A, $crnd[c]$, $i$) to $a$, where, $\forall a, a \in Q_a$
11. **If** $((leader = true) \wedge (i > I))$ **then**
12.     $v \leftarrow id : (\exists id, id \in request, request \in req\_set$ and this coordinator has not sent $id$ in any PHASE 2A message)
13.     $cval[c] \leftarrow v$, $crnd[c] \leftarrow lsn$
14.     write ($crnd[c]$, $cval[c]$, $i$) to stable storage
15.     send (PHASE 2A, $crnd[c]$, $cval[c]$, $i$) to successor
16. **upon** (electing a new leader)
17.     **if** $((leader = true) \wedge (i \leq I))$ **then**
18.         $crnd[c] \leftarrow lsn$
19.         send ( PHASE 1A, $crnd[c]$, $i$) to $a$, where, $\forall a, a \in Q_a$
20. **upon** receiving (PHASE 1B, $rnd[a], vrnd[a], vval[a], i$) from any $Q_a$ such that $(crnd[c] = rnd[a])$, where, $\forall a, a \in Q_a$
21.     $k \leftarrow \max(vrnd[a] : \forall a, a \in Q_a)$
22.     $v \leftarrow id : (\exists id, id \in request, request \in req\_set$ and this coordinator has not sent $id$ in any PHASE 2A message)
23.     **if** ($k = 0$) **then** $cval[c] \leftarrow v$
24.     **else**
25.         let $V$ be a set of value of $vval[a]$ received in PHASE 1B messages from $\forall a, a \in Q_a$ such that corresponding $vrnd[a] = k$, $V$ contains a single element, let $u \in V$.
26.         $cval[c] \leftarrow u$
27.     write ($crnd[c], cval[c], i$) to stable storage
28.     send (PHASE 2A, $crnd[c], cval[c], i$) to successor
29. **upon** receiving (denial)
30.     $leader \leftarrow false$
31. **upon** receiving (PHASE 2B, $crnd[c]$, $cval[c]$, $i$) from any acceptor
32.     multicast (LEARNED, $cval[c]$, $i$) to all acceptors and learners
33. /* Task for any acceptor  a */
34. **upon** (startup)
35.     read $(rnd[a], vrnd[a], vval[a], sn)$ from stable storage for the instance number $i$, if exist
36.     **if** $(vrnd[a] = lsn)$ **then**
37.         **if** $(sn < (|Q_a| - 1))$ **then** send (PHASE 2A & 2B, $crnd[c]$, $cval[c]$, $sn$, $i$) to successor
38.         **else** send (PHASE 2B, $crnd[c]$, $cval[c]$, $i$) to coordinator $c$
39. **upon** receiving ( PHASE 1A, $crnd[c]$, $i$) from any coordinator $c$
40.     **if** $(crnd[c] > rnd[a])$ **then**

41.        $rnd[a] \leftarrow crnd[c]$
42.       send (PHASE 1B, $rnd[a]$, $vrnd[a]$, $vval[a]$, $i$) to the coordinator $c$
43.    **else if** $(crnd[c] < rnd[a])$ **then** send $(denial)$ to the coordinator $c$
44. **upon** receiving (PHASE 2A, $crnd[c]$, $cval[c]$, $i$) from coordinator $c$ or upon receiving (PHASE 2A & 2B, $crnd[c]$, $cval[c]$, $sn$, $i$) any acceptor $a$
45.   $sn \leftarrow sn + 1$
46.   **If** $((crnd[c] \geq rnd[a]) \wedge (crnd[c] \neq vrnd[a]))$ **then**
47.      $rnd[a] \leftarrow crnd[c]$
48.      $vrnd[a] \leftarrow crnd[c]$
49.      $vval[a] \leftarrow cval[c]$
50.     **If** $(sn < (|Q_a| - 1))$ **then**
51.        write ($rnd[a]$, $vrnd[a]$, $vval[a]$, $sn$, $i$) to stable storage
52.        send (PHASE 2A & 2B, $crnd[c]$, $cval[c]$, $sn$, $i$) to successor
53.     **else**
54.        write ($rnd[a]$, $vrnd[a]$, $vval[a]$, $sn$, $i$) to stable storage
55.        send (PHASE 2B, $crnd[c]$, $cval[c]$, $i$) to coordinator $c$
56. **upon** receiving (LEARNED, $cval[c]$, $i$) from coordinator $c$
57.    **if** (learner is not available on this site ) **then**
58.      $learned\_set \leftarrow (cval[c], i)$
59.     **If** (*request* was received from a proposer) **then** send (*id*) to the corresponding proposer such that $id \in request$
60.     Close HT-Ring Paxos(*i*)
61. /* *Task for any learner l* */
62. **upon** receiving (LEARNED, $cval[c]$, $i$) from any coordinator $c$
63.    **if** $(\exists request : request \in req\_set \wedge cval[c] \in request)$ **then**
64.      Execute *request* after completing the execution of previous (*i* - 1) *requests*
65.    **else**
66.     $id \leftarrow cval[c]$
67.    send (*id*) to any randomly chosen coordinator
68.     Repeat from step 67, after every $\Delta$ time, upon not receiving *request* such that $cval[c] \in request$
69.     Execute *request* after completing the execution of previous (*i* - 1) *requests*
70.    $learned\_set \leftarrow (cval[c], i)$ //optional
71.    **If** (*request* was received from a proposer) **then** send (*id*) to the corresponding proposer such that $id \in request$
72.    Close HT-Ring Paxos(*i*)

---

### 3.3 Optimizations of HT-Ring Paxos

We introduce a few optimizations in HT-Ring Paxos, most of which have been described previously in the literature. Under high load conditions, any broadcaster/acceptor may wait for more client requests for making a batch of client requests and then creates a batch *id*, now it multicasts <batch, batch *id*>. Leader can achieve consensus for these batch *ids*. At the leader, Phase 2 is executed for a batch of proposed *ids*, and not for a single *id*; one consensus instance can be started before the previous one has finished. Placing any suitable quorum system [31] [32] for the construction of a ring can reduce the number of communication steps to reach a decision and may increase fault tolerance but at the cost of availability (for the construction of a ring in HT-Ring Paxos, we have earlier considered a majority quorum systems [7] for the higher availability). Finally, leader can club various multicast messages (if they are concurrently available to send) into a single one and then multicast.

Moreover, instead of having a fixed ring of acceptors, we can choose an alternative approach. We can assign each acceptor a unique natural number $j$ in a sequence, starting from one. Any acceptor $((j+d) \mod n)$ will be the successor of acceptor $j$, where, initially, $d$ is a natural number equal to one. Any acceptor (including leader) will always acknowledge to the sender acceptor (including leader). After a certain timeout, if acceptor does not receive an acknowledgment then it increases $d$ by one and again sends the message to the successor. This approach does not require a view change on every failure of an acceptor in the ring. However leader will receive an extra acknowledgment from the first acceptor of the ring and will send an extra acknowledgment to the last acceptor of the ring under normal operations.

### 3.4 Safety

#### 3.4.1 Safety Criteria

For the safety of any protocol that implements state machine replication, no two learners can learn the values in different order despite any number of failures (in our case, non-Byzantine failures).

#### 3.4.2 Proof of Safety (Sketch)

Our proposed protocol fulfills the safety requirement by adopting the following provisions,

*Nontriviality:* only proposed values can be learned.

As per the proposed protocol, leader proposes only *ids* of client requests or batch of client requests and after receiving Phase 2B message from an acceptor leader decides only on these proposed *ids*. Thereby learners learn these proposed *ids* i.e. learners can only learn proposed values.

*Consistency:* learners can learn the client requests only in same sequence.

Since, HT-Ring Paxos is derived from the classical Paxos. After dissemination of client requests in HT-Ring Paxos, ordering of *ids* is basically based on classical Paxos, the only difference here is how the messages flow among the various agents. Learners can determine the order of client requests by the help of the learned *ids*. Classical Paxos is a well-proven theory of literature that guarantees safety; therefore, no two learners can learn the values (*client requests*) in different order. ∎

## 3.5 Progress

Under certain assumptions, HT-Ring Paxos ensures that all available learners will surely learn the client *request*. Moreover, protocol also ensures that if client does not crash for an enough time then client will definitely receive a reply for their *request*.

### 3.5.1 Requirements for ensuring progress

At least $\lfloor n/2 \rfloor + 1$ acceptors out of total *n* and one learner are always required to remain non-faulty for the progress of the proposed protocol (these requirements are only for ensuring progress, safety does not require these conditions).

### 3.5.2 Proof of Progress (sketch)

As per the system model, a majority of acceptors (coordinators) will always remain non-faulty, i.e. any request proposed by any client will definitely reach to the leader either directly or indirectly through other coordinators. Leader launches the instances of HT-Ring Paxos for the purpose of ordering of client requests through their *ids*; these instances uses the messages and process these messages as per the concept of classical Paxos, only difference here is that how these messages reach to their destination. Since, the classical Paxos is a well-defined theory of literature that ensures progress. Therefore, we can say that under aforementioned specific requirements, HT-Ring Paxos ensures progress. ∎

## 4. Comparative Analysis

Throughput of any system can be increased by increasing the processing power of computers and increasing the bandwidth of communication network. This solution may not be suitable for increasing throughput every time for either technological and/or economic reasons.

Alternatively, we can adopt a more scalable and throughput efficient state machine replication protocol, i.e., a state machine replication protocol that requires comparatively less computation at individual computers and less data communication at individual LANs/ individual computers. In addition, more scalable state machine replication protocol allows us to increase throughput by increasing more computers and more LANs. However, after a certain limit, it cannot scaled up because of coordination overload; in fact, it may start reducing the throughput. This limit depends on the protocol that we use.

Compare to earlier versions of Paxos (like classical Paxos, fast Paxos or generalized Paxos), new variants of Paxos (like ring Paxos, multi-ring Paxos, S-Paxos and HT-Paxos) increase the scalability and throughput by reducing the processing and bandwidth requirements at the busiest computing node. Now, we start comparing the processing and bandwidth requirements of various Paxos protocols that affects system scalability and throughput.

## 4.1 Processing requirements

In general, processing requirements reduce at any individual computing node, if it requires to response or process a less number of messages for state machine replication protocols. Therefore, we require analyzing the total number of messages required at the busiest node for some given number of client requests.

Moreover, Data communication also requires some processing at any individual computing node. Higher data communication at any individual computing node also requires higher processing requirement. We will discuss this requirement in the next *bandwidth requirements* section.

### *4.1.1  Comparative message analysis*

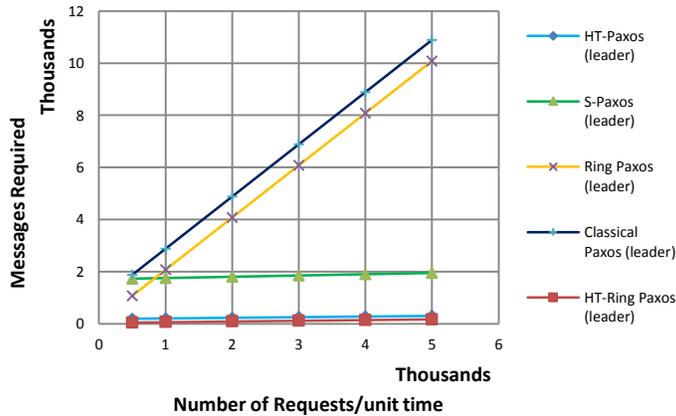

**Fig. 2** Comparison among mentioned variants of Paxos for the messages requirements at the busiest computing nodes, where total number of acceptors (acceptors/disseminators in case of S-Paxos and HT-Paxos) = 40.

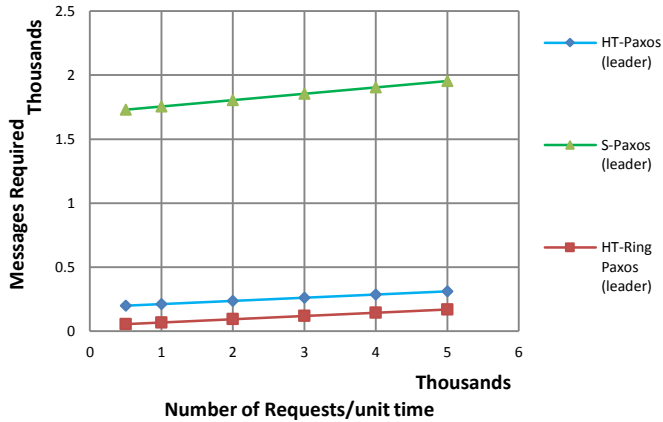

**Fig. 3** Comparison among mentioned variants of Paxos for the messages requirements at the busiest computing nodes, where, total forty acceptors (disseminators/acceptors in case of S-Paxos and HT-Paxos) = 40.

As shown in Fig. 2, the very high number of messages in classical Paxos and ring Paxos are because of the fact that all client communication passes through the leader. S-Paxos, HT-Paxos and HT-Ring Paxos decentralize the client communication. As shown in Fig. 3, Message advantage of HT-Paxos over S-Paxos is because of the fact that in S-Paxos, every disseminator is required to send an acknowledgement message to every other disseminator, in HT-Paxos acknowledgement message goes only to the sender disseminator. Moreover, HT-Ring Paxos have the advantage over S-Paxos and HT-Paxos because of the ring of the acceptors and it does not require at all the acknowledgement messages as these were used in the S-Paxos and HT-Paxos for the purpose of stabilizing the requests.

## 4.2 Bandwidth requirements

Bandwidth requirement of any communication network depends upon the size of data (requests) and number of messages required to communicate by any computer. If any protocol requires more messages than due to message overhead, more data will pass through the communication network, hence will requires higher bandwidth.

In any data center, bottleneck may be the bandwidth of communication network, in such a case, either we replace the lower bandwidth LAN with higher bandwidth LAN or adopt multiple LANs. First option may not be convenient for either technological or economic reasons. In data centers, we do not require big data cables; therefore, economically it may not a big issue in any large data center.

However, if bottleneck is network sub system of any computing node that handles data communication (because of handling more data) then replacement of computing node with higher processing power computing node may really be a big issue. Therefore, it is important to analyze the bandwidth requirements of individual computing nodes of the various variant of Paxos.

For the analysis of bandwidth requirements of individual computing nodes, we are considering the message overhead of 128 bytes (as IP packet header, Ethernet frame preamble, gap, header, footer, other network protocols like ARP etc and variables used in the protocol create overheads). Bigger message overhead will be in the favor of our proposed protocol, because our proposed protocol requires fewer messages as mentioned above,

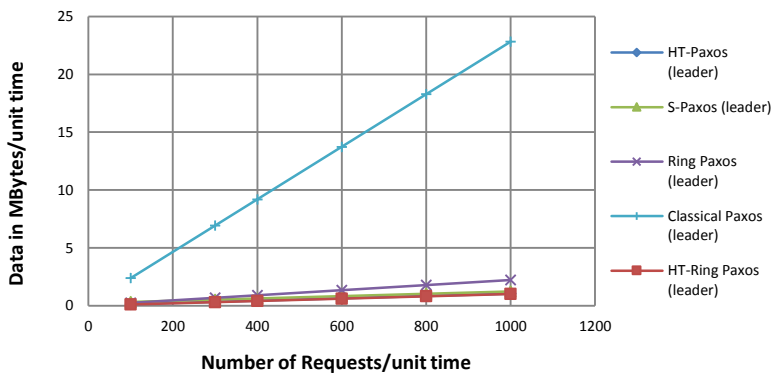

**Fig. 4** Comparison of bandwidth requirements at the mentioned computing nodes of the various mentioned variant of Paxos, where, total number of acceptors (acceptors/disseminators in case of S-Paxos and HT-Paxos) = 40, and data size of request = 1k bytes

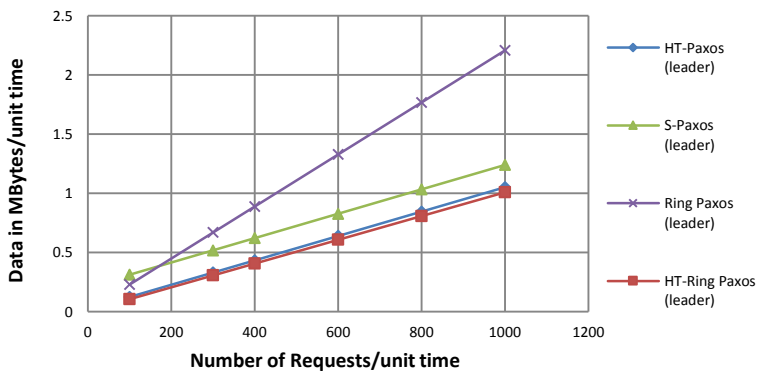

**Fig. 5** Comparison of bandwidth requirements at the mentioned computing nodes of the various mentioned variant of Paxos, where, total number of acceptors (acceptors/disseminators in case of S-Paxos and HT-Paxos) = 40, and data size of request = 1k bytes.

As shown in Fig. 4, leader of classical Paxos handles extremely large amount of data, because this protocol achieves consensus on request (or batch of requests) rather than request *id* or batch *id*. Other variants of Paxos for high throughput achieves consensus on request *id* or batch *id*.

As shown in Fig. 5, leader of ring Paxos handles large amount of data as compare to other variants of high throughput Paxos, because leader of ring Paxos handles all client communications. In case of fewer requests, ring Paxos performs better than S-Paxos, because of comparatively large number of acknowledgement messages at the leader (also at other disseminators) of S-Paxos.

As shown in Fig. 5, leader of HT-Paxos handles less data compare to leader of classical Paxos, ring Paxos and S-Paxos because of decentralized client communications like S-Paxos; however, because of fewer acknowledgement messages compare to S-Paxos, HT-Paxos handles even less data compare to S-Paxos.

As shown in Fig. 5, leader of HT-Ring Paxos handles less data compare to all mentioned variants of Paxos, because of decentralized client communications like S-Paxos and HT-Paxos. However, because of the ring of the acceptors and it no acknowledgement messages the HT-Paxos handles least data among all variants of Paxos.

## 4.3 Latency

When client directly send their request to the leader, HT-Ring Paxos takes ($m$ +2) message delays in the best case for learning the client request, where, $m$ represents a total number of acceptors in the ring. Otherwise, if client sends their request to any broadcaster except leader then it takes ($m$ +3) message delays in the best case. If client sends their request to any coordinator except broadcasters then it takes ($m$ +3) message delays in the best case, if this coordinator sends request directly to the leader in the next step. Otherwise, it takes ($m$ +4) message delays in the best case. Moreover, Ring Paxos take ($m$ +2) message delays in the best case.

HT-Paxos and S-Paxos take six message delays in the best case (for message-optimized version of classical Paxos in the ordering layer). While, classical Paxos takes four message delays in message-optimized version and three message delays otherwise in the best case. Moreover, fast Paxos and generalized Paxos take only two message delays in the best case.

## 4.4 Response time

When client directly send their request to the leader, HT-Ring Paxos takes ($m$ +2) message delays in the best case for responding to the client request, where, $m$ represents a total number of acceptors in the ring. Otherwise, if client sends their request to any broadcaster except leader then it takes ($m$ +4) message delays in the best case. If client sends their request to any coordinator except broadcasters then it takes ($m$ +4) message delays in the best case, if this coordinator sends request directly to the leader in the next step. Otherwise, it takes ($m$ +5) message delays in the best case. Moreover, Ring Paxos takes ($m$ +2) message delays in the best case.

HT-Paxos takes four message delays for responding to the client *request* in the best case. S-Paxos take six message delays in the best case. Moreover, classical Paxos takes only four message delays (for message-optimized version of classical Paxos).

## 4.5 Other Related Work

Zab [24] is a variant of the Paxos, basically designed for the yahoo's Zookeeper coordination service (a primary-backup data replication system). In zookeeper, any client sends their request to any replica (either leader or follower). Follower replica forwards all update requests to the primary replica for taking the services of Zab. Zab is a centralized protocol, where primary replica disseminates the update requests to all other replicas and the leader that generally is on the same primary replica site works for ensuring a proper order of the requests. Because of the centralized approach of the Zab, in any large clustered data centers bottleneck may be the resources of the leader's site (or primacy's site as Zab considers both on the same site). Therefore, under very high workload conditions throughput and scalability will obviously be less in any large clustered data center.

Mencius [18] works on a moving sequencer approach [9] to prevent the leader from becoming the bottleneck. It partitions the sequence of instances of the protocol among all replicas and each replica becomes a leader of an instance in a round-robin fashion. After failure of any replica, Mencius reconfigures the system to exclude all failed replicas. Moreover, even in the case of failure free execution, leader of Mencius disseminates as well as orders all the available client requests. Under heavy load environment, leader of Mencius will handle more number of messages as well as more data as compared HT- Ring Paxos. However, Mencius was designed for optimizing state-machine replication protocol for WAN environment. Contrary to this HT-Ring Paxos is for clustered environment.

LCR [22] is a high throughput state-machine replication protocol base on virtual synchrony model [20] instead of Paxos. LCR arranges replicas in a logical ring and uses vector clocks for message ordering. LCR is a very much decentralized protocol; thereby it utilizes all available system resources properly. However, latency and response time increases linearly with the number of replicas in the ring. In large clustered data centers as we are considering for, this number could be very significant. Although, LCR has slightly better bandwidth efficiency, Moreover, LCR requires reconfiguration of the system after every failure for ensuring progress and *perfect failure detection* is required i.e. erroneously considering a replica to have crashed is not tolerated, it implies stronger synchrony assumptions.

Multi-Ring Paxos [27] uses the concept of State partitioning [8] for achieve scalability. Multi-Ring Paxos have various logical groups. Each logical group has an instance of ring Paxos. Learner may subscribe to any one or more logical groups. If a learner subscribes to multiple logical groups then Multi-Ring Paxos uses a deterministic procedure to merge messages coming from different instances of ring Paxos. However, proposed HT-Ring Paxos can also adopt the concept of state partitioning as broadcasters can multicast the client request to only interested learners.

## 5. Conclusion And Future Work

Paxos based protocols are very prominent for replica control. Earlier versions of Paxos (like classical Paxos, fast Paxos or generalized Paxos) were more focused on *fault tolerance* and *latency* but *throughput* was comparatively low. However, in current scenario, *throughput* requirement is increasing significantly. HT-Ring Paxos is a variant of classical Paxos designed for achieving significantly high *throughput* and *scalability*. Moreover, it is best suitable for large clustered data centers and achieves the best *throughput* and *scalability* among all variants of Paxos.

Throughput may be limited because of (i) processing power of CPU or (ii) data handling capacity of network sub system of any computing node or (iii) bandwidth of communication networks. Since, in clustered data centers, computing resources are generally more costly than data cables. Therefore, high throughput replica control protocols should avoid bottleneck of CPU and network subsystems through less computing requirements of CPU and less bandwidth requirements at any individual computing node. HT-Ring Paxos achieves all these goals very significantly for improving throughput and scalability.

However, latency and response time of the HT-Ring Paxos as compared to other high throughput state-machine replication protocols are as high as of Ring Paxos in the best case otherwise slightly higher. Moreover, HT-Ring Paxos achieves the same fault tolerance as classical Paxos.

As future work, we plan to apply our technique to Byzantine faults, and will optimize HT-Ring Paxos for WAN.

# References

[1] Leslie Lamport, "Time, clocks, and the ordering of events in a distributed system," *Communications of the ACM*, 21(7):558–565, 1978.
[2] M. Pease, R. Shostak, L. Lamport, "Reaching agreement in the presence of faults," Journal of ACM, 228-234 (1979).
[3] Leslie Lamport, "The part-time parliament," *ACM Transactions on Computer Systems*, 16(2):133–169, May 1998.
[4] Leslie Lamport. Paxos made simple. *ACM SIGACT News (Distributed Computing Column)*, 32(4):18{25, December 2001.
[5] Leslie Lamport, "Fast Paxos," Distributed Computing, vol. 19, no. 2, pp. 79–103, 2006.
[6] F. Schneider, "Implementing fault-tolerant services using the state machine approach: A tutorial," ACM Computing Survey vol. 22, no. 4, pp. 299–319, December 1990.
[7] R.H. Thomas," A majority consensus approach to concurrency control for multiple copy database," ACM Trans. Database Systems 4 (2) (1979) 180-209.
[8] J. Gray, P. Helland, P. O'Neil, and D. Shasha, "The dangers of replication and a solution," in *SIGMOD '96*, 1996.
[9] X. Defago, A. Schiper, and P. Urban, "Total order broadcast and multicast algorithms: Taxonomy and survey," *ACM Computing Surveys*, vol. 36, p. 2004, 2004.
[10] Leslie Lamport. The implementation of reliable distributed multiprocess systems. Computer Networks, 2:95–114, 1978.
[11] Leslie Lamport, "Generalized Consensus and Paxos" Microsoft Research Technical Report MSR-TR-2005-33 (2005).
[12] Francisco Brasileiro, Fab¶³ola Greve, Achour Mostefaoui, and Michel Raynal. Consensus in one communication step. In V. Malyshkin, editor, *Parallel Computing Technologies (6th International* Conference*, PaCT 2001)*, volume 2127 of *Lecture Notes in Computer* Science, pages 42{50. Springer-Verlag, 2001.
[13] Jean-Philippe Martin and Lorenzo Alvisi. Fast byzantine consensus. In *Proceedings of the International Conference on* Dependable *Systems and Networks (DSN 2005)*, Yokohama, June 2005. IEEE Computer Society.
[14] Michael Burrows. The Chubby lock service for loosely-coupled distributed systems. In Proc. 7th USENIX OSDI, Seattle, WA, November 2006.
[15] R. Ekwall and A. Schiper, "Solving atomic broadcast with indirect consensus," in *DSN'06*, 2006, pp. 156–165.
[16] Tushar Chandra, Robert Griesemer, Joshua Redstone "Paxos Made Live – An Engineering Perspective" PODC '07: 26th ACM Symposium on Principles of Distributed Computing (2007).
[17] J. Kirsch and Y. Amir, "Paxos for system builders," Dept. of CS, Johns Hopkins University, Tech. Rep., 2008.
[18] Y. Mao, F. P. Junqueira, and K. Marzullo, "Mencius: building efficient replicated state machines for wans," in *OSDI'08*, 2008, pp. 369–384.
[19] Vigfusson, Ymir, Hussam Abu-Libdeh, Mahesh Balakrishnan, Ken Birman, Robert Burgess, Gregory Chockler, Haoyuan Li, and Yoav Tock. "Dr. multicast: Rx for data center communication scalability." In *Proceedings of the 5th European conference on Computer* systems, pp. 349-362. ACM, 2010.
[20] K.P. Birman. A history of the Virtual Synchrony replication model. In B. Charron-Bost, F. Pedone, and A. Schiper, editors, Replication: theory and Practice, volume 5959 of Lecture Notes in Computer Science, chapter 6, pages 91–120. Springer-Verlag, 2010.
[21] L. Lamport, D. Malkhi, and L. Zhou, "Reconfiguring a state machine,"*SIGACT News*, vol. 41, pp. 63–73, 2010.
[22] R. Guerraoui, R. R. Levy, B. Pochon, and V. Qu´ema, "Throughput optimal total order broadcast for cluster environments," *ACM Trans. Comput. Syst.*, vol. 28, pp. 5:1–5:32, 2010.
[23] P. J. Marandi, M. Primi, N. Schiper, and F. Pedone, "Ring Paxos: A high-throughput atomic broadcast protocol," in *DSN'10*, 2010, pp. 527–536.
[24] F. P. Junqueira, B. C. Reed, and M. Serafini, "Zab: High-performance broadcast for primary-backup systems," in *DSN'11*, 2011, pp. 245–256.
[25] Baker, Jason, Chris Bond, James Corbett, J. J. Furman, Andrey Khorlin, James Larson, Jean-Michel Léon, Yawei Li, Alexander Lloyd, and Vadim Yushprakh. "Megastore: Providing Scalable, Highly Available Storage for Interactive Services." In *CIDR*, vol. 11, pp. 223-234. 2011.
[26] P. Marandi, M. Primi, and F. Pedone, "High performance state-machine replication," in *Dependable Systems Networks (DSN), 2011 IEEE/IFIP 41st International Conference on*, june 2011, pp. 454 –465.
[27] Marandi, Parisa Jalili, Marco Primi, and Fernando Pedone. "Multi-Ring Paxos." In *Dependable Systems and Networks (DSN), 2012 42nd Annual IEEE/IFIP International Conference on*, pp. 1-12. IEEE, 2012.
[28] N. Santos and A. Schiper, "Tuning Paxos for high-throughput with batching and pipelining," in *13th International Conference on Distributed Computing and Networking (ICDCN 2012)*, Jan. 2012.
[29] M. Biely, Z. Milosevic, N. Santos, and A. Schiper. "S-Paxos: Offloading the Leader for High Throughput State Machine Replication." (2012).
[30] L. Lamport and M. Massa. Cheap Paxos. In International Conference on Dependable Systems and Networks (DSN), pages 307–314, 2004.
[31] Kumar, Vinit and Ajay Agarwal, "Generalized grid quorum consensus for replica control protocol," In proceedings of IEEE CICN-2011, pp 395-400.
[32] Kumar, Vinit, and Ajay Agarwal. "An Efficient Read Dominant Data Replication Protocol under Serial Isolation using Quorum Consensus Approach." *arXiv preprint arXiv:1406.7423* (2014).
[33] Kumar, Vinit, and Ajay Agarwal. "HT-Paxos: High Throughput State-Machine Replication Protocol for Large Clustered Data Centers." *arXiv preprint arXiv:1407.1237* (2014).